\documentclass[a4paper,10pt]{article}
\usepackage{enumerate}
\usepackage{amssymb, amsmath, amsfonts, amsthm}
\usepackage[utf8]{inputenc} 
\usepackage{float}
\usepackage{graphicx}
\usepackage{flushend}
\usepackage{multicol}
\usepackage[left=1.7cm,top=2.2cm,right=1.7cm,bottom=2.5cm]{geometry}

\usepackage{multirow}
\usepackage{slashed}
\usepackage{mathrsfs}
\usepackage{soul}
\usepackage{color}
\usepackage{ wasysym }
\newcommand{\la}{\langle}
\newcommand{\ra}{\rangle}
\newcommand{\s}{\slashed}
\newcommand{\g}{\gamma}
\newcommand{\ba}{\begin{align}}
\newcommand{\ea}{\end{align}}
\newcommand{\be}{\begin{equation}}
\newcommand{\ee}{\end{equation}}

\begin{document}

\begin{center}
\Large{\textbf{Helicity Amplitudes for massive gravitinos in $\mathcal{N}=1$ Supergravity}}
\normalsize
\\ \vspace{0.5cm}
J. Lorenzo Diaz-Cruz and Bryan O.~Larios  \\
\it{\normalsize Facultad de Ciencias F\'isico-Matem\'aticas, Benem\'erita Universidad Aut\'onoma de Puebla, 
Av. San Claudio y 18 Sur, C. U. 72570 Puebla, M\'exico}
\end{center}

\smallskip

\noindent

\large
\noindent

\smallskip
\normalsize
\indent \textbf{Abstract.} 
We develop the formal tools needed to construct helicity amplitudes for massive gravitino in $\mathcal{N}=1$ SUGRA. 
We start by considering the helicity states for massive spin-3/2 particles, 
which involves the solutions of Rarita-Schwinger equation. These solutions are written
using the modern  spinor bra-ket notation  and are used then to derive the interactions of gravitino with
matter and gauge supermultiplets within $\mathcal{N}=1$ Supergravity. 
The corresponding interactions of  goldstinos are discussed too, relying on the goldstino-gravitino equivalence theorem. 


\section{Introduction}\label{introduction}

Great progress has been made in recent years to understand the amplitudes in gauge theories, including gravitation 
and Yang-Mills fields \cite{PT}-\cite{KLT}. Impressive results for multi-leg amplitudes in the massless case have been derived \cite{BCF, BCFW},
which allow to evaluate multiparticle final states for maximal helicity violating amplitudes.

Some of the results have been derived for $\mathcal{N}=4$ Super Yang Mills theory, which turns out to provide some
regularities that make it to look as "The simplest Quantum Field Theory" \cite{Nima3}. Exploring whether the local SUSY theory also knows as Supergravity, retains some of these properties would be quite  interesting.  One would like to have similar progress for the massive case, both from its formal relevance as well as the 
phenomenological implications namely, colliders like LHC are aimed to study massive states, such as the top quark, W, Z, Higgs, which have a   mass that is not negligible as compared with the CM energy. Ideally, if possible we would like to understand the mass effects as perturbations from the massless case. It is possible that we can learn about this by studying specific cases.

With this aim we are interested in studying the application of helicity methods to treat the amplitudes involving the massive
gravitino, which appears as the superpartner of the graviton in minimal $\mathcal{N}=1$ Supergravity. Studying the gravitino
has a relevance in its own in particle physics and cosmology, partly because when the Minimal SUSY extension 
of the standard model is embedded within SUGRA, the SUSY spectrum include the gravitino as the lightest 
SUSY particle, and therefore it could become a candidate for dark matter \cite{le}. In fact supersymmetric extension of the standard model of particle physics have been thoroughly   studied theoretically and its effects and predictions have been searched at low and high energies.

Studying the gravitino properties and its implications for both collider physics 
and the early universe, require the evaluation of many processes which could be quite
involved due to the form of the spin-3/2 propagators and wave-functions for external legs.
Some simplifications can arise for very light gravitinos, where one can rely on the equivalence
theorem and replace the longitudinal components (spin-1/2) of the gravitino by the goldstino
coming from the Chiral superfield appearing in the super-Higgs mechanism \cite{gge,fayet1,fayet2,fayet3,fayet4,fayet5}.

We have already considered some aspects of gravitino phenomenology \cite{us1}, in particular we studied the stop decay 
$\tilde{t}\to t\,W\,\tilde{\Psi}_{\mu}$  \cite{le}, which already shows some complications. We would like to work with a formalism 
based on helicity methods to deal with such decays, as well as other process appearing in gravitino phenomenology. 
Some calculations dealing with gravitino were presented some time ago \cite{novaes, roy1, roy2}; more
modern methods have been incorporated into general programs such as Madgraph \cite{madgraph}. 
However, these methods have still some limitations, such as give only numerical outputs and not all vertices of the general SUGRA  are included into the program.

In general, the incorporation of the massive case is not treated
with full generality in the literature, which is one of the goals of this paper.
Thus, we shall present  the implementation of the Feynman rules for gravitino with an appropriate notation, 
which allows to reduce huge calculation which are very difficult to compute analytically using the traditional 
approach. In the language of Hunters, we want to show the guts and not only the skin of cross section and amplitudes.

After this introductory Section \ref{introduction}, let us present the organization of our paper. In Section \ref{helicitys_pinor_formalism}  we shall discuss the 
solutions of the Rarita-Schwinger equation appropriate to be implemented with helicity methods. The helicity amplitudes with the full spin-3/2 gravitino and with the goldstino approximation are presented in Section \ref{helicity_amplitudes}, finally  Section \ref{comparison_amplitudes} includes some applications were we compared the helicity amplitudes for the 2-body neutralino decay with gravitino and goldstino in the final state
as well as the reactions: $e^{+}\,e^{-}\to GG$. Some details and conventions are left to the appendices.


\section{Helicity Spinor Formalism for spin-$3/2$ gravitino field}\label{helicitys_pinor_formalism}
In order to compute Scattering Amplitudes (SA)  with  spin-3/2 gravitino field in the final state, 
we shall use the marvelous advantages that the Spinor Helicity Formalism (SHF) \cite{schwartz, srednicki, elvang, lorenzo} provides to handle perturbative calculation in quantum fields theories. In principle  we want to compute SA considering massive particles, hence it will be necessary to use to the Light Cone Decomposition (LCD) \cite{boels, weinzierl, spinorsextras, BOL} which helps for expressing massive momenta in terms of massless ones. In Appendix A just by completeness  we review   some basics properties of the massless SHF that will also be useful for the massive extension.  

The Rarita-Schwinger equation \cite{rarita,auvil,takeo} is equivalent to the following set of equations
\begin{align}
\g_{\mu}\tilde{\Psi}^{\mu}_{\lambda_p}(p)&=0,\label{eq1:irreduciblecondition}\\
p_{\mu}\tilde{\Psi}^{\mu}_{\lambda_p}(p)&=0,\label{eq2:irreduciblecondition}\\
(\s{p}-\tilde{m})\tilde{\Psi}^{\mu}_{\lambda_p}(p)&=0.\label{eq3:irreduciblecondition}
\end{align}

The 4 polarization states of the gravitino in the momentum space (in terms of spin-1 and spin-1/2 components) that fulfill 
these equations are as follows
\begin{align}
\tilde{\Psi}_{++}^{\mu}(p)&=\epsilon_{+}^{\mu}(p)u_+(p),\label{eq:gravitinostate01}\\
\tilde{\Psi}_{--}^{\mu}(p)&=\epsilon_{-}^{\mu}(p)u_-(p),\label{eq:gravitinostate02}\\
\tilde{\Psi}_{+}^{\mu}(p)&=\sqrt{\frac{2}{3}}\epsilon_{0}^{\mu}(p)u_+(p)+\frac{1}{\sqrt{3}}\epsilon_{+}^{\mu}(p)u_-(p),\label{eq:gravitinostate03}\\
\tilde{\Psi}_{-}^{\mu}(p)&=\sqrt{\frac{2}{3}}\epsilon_{0}^{\mu}(p)u_-(p)+\frac{1}{\sqrt{3}}\epsilon_{-}^{\mu}(p)u_+(p),\label{eq:gravitinostate04}
\end{align}

It is known in literature how  to express the polarization vectors $\epsilon_{\pm}^{\mu}(p),\,\epsilon_{0}^{\mu}(p)$  as well as the massive Dirac spinors $u_{\pm}(p)$ in terms of the modern and powerful bra-kets notation \cite{dittmaier} (see Ref.~\cite{byo} for a detailed  review of massive SHF and its applications to QED, EWSM and Physics Beyond the Standard Model). It is straightforward to express  the 4 gravitino states in this bra-kets notation, this are as follows  
\begin{align}
\tilde{\Psi}^{\mu}_{++}(p)&=\frac{\la r|\g^{\mu}|q]}{\sqrt{2}[rq]}\left(|r\ra+\tilde{m}\frac{|q]}{[rq]}\right),\\
\tilde{\Psi}^{\mu}_{--}(p)&=\frac{\la q|\g^{\mu}|r]}{\sqrt{2}\la rq\ra}\left(|r]+\tilde{m}\frac{|q\ra}{\la rq\ra}\right),\\
\tilde{\Psi}^{\mu}_{-}(p)&=\sqrt{\frac{2}{3}}\left(\frac{r^{\mu}}{\tilde{m}}-\tilde{m}\frac{q^{\mu}}{s_{qr}}\right)\left(|r]+\tilde{m}\frac{|q\ra}{\la rq\ra}\right)+\frac{1}{\sqrt{3}}\frac{\la q|\g^{\mu}|r]}{\sqrt{2}\la rq\ra}\left(|r\ra+\tilde{m}\frac{|q]}{[rq]}\right),\\
\tilde{\Psi}^{\mu}_{+}(p)&=\sqrt{\frac{2}{3}}\left(\frac{r^{\mu}}{\tilde{m}}-\tilde{m}\frac{q^{\mu}}{s_{qr}}\right)\left(|r\ra+\tilde{m}\frac{|q]}{[rq]}\right)+\frac{1}{\sqrt{3}}\frac{\la r|\g^{\mu}|q]}{\sqrt{2}[rq]}\left(|r]+\tilde{m}\frac{|q\ra}{\la rq\ra}\right),
\end{align}
where the 4-momenta $r^{\mu}$ and $p^{\mu}$ are massless, and the Mandelstam variable is  $s_{qr}=-(q+r)^2=-2q\cdot r$. Before go ahed one has to check if the 4 gravitino states in this new notation fulfill the equations (\ref{eq1:irreduciblecondition})-(\ref{eq3:irreduciblecondition}) as well as the normalization condition 
\begin{equation}\label{eq01:normalizationcondition}
\bar{\tilde{\Psi}}_{\lambda_{1}\mu}(p)\tilde{\Psi}_{\lambda_2}^{\mu}(p)=2\tilde{m}\delta_{\lambda_1\lambda_2}.
\end{equation}
It shall be useful rearrange  the 4 gravitino states as an expansion on the gravitino mass $\tilde{m}$  
\begin{align}
\tilde{\Psi}_{++}(p)&=\beta_1^{\mu}|r\rangle+\beta_2^{\mu}|q]\tilde{m},\label{eq:gravitinobasis1}\\
\tilde{\Psi}_{--}(p)&=-\beta_1^{*\mu}|r]+\beta_2^{*\mu}|q\rangle\tilde{m},\\
\tilde{\Psi}_-^{\mu}(p)&=\beta_3^{\mu}|r]+(\beta_4^{\mu}|q\rangle+\beta_5^{\mu}|r\rangle)\tilde{m}+(\beta_6^{\mu}|r]+\beta_7^{\mu}|q])\tilde{m}^2+\beta_8^{\mu}|q\rangle\tilde{m}^3,\\
\tilde{\Psi}_+^{\mu}(p)&=\beta_3^{*\mu}|r\ra-(\beta_4^{*\mu}|q]+\beta_5^{*\mu}|r])\tilde{m}+(\beta_6^{*\mu}|r\ra+\beta_7^{*\mu}|q\ra)\tilde{m}^2-\beta_8^{*\mu}|q]\tilde{m}^3,\label{eq:gravitinobasis4}
\end{align}
the gravitino mass $\tilde{m}$ is directly connected with the the SUSY breaking energy scale $F$ as $\tilde{m}=\frac{F}{\sqrt{3}M}$, where $M$ is the Plank mass. 
we have defined all the $\beta_i^{\mu}\,\forall\,i=1\dotsm8$ in the next table:\\
\begin{table}[H]
\begin{center}
  \begin{tabular}{  || c | c |c || }
    \hline \hline
  $i$  & $\beta_i^{\mu}$ & $\beta_i^{*\mu}$ \\ \hline
   \hline\hline
    $1$ & $\frac{\la qr\ra \la r |\g^{\mu}|q]}{\sqrt{2}s_{qr}}$ & $\frac{ [rq] \la q |\g^{\mu}|r]}{\sqrt{2}s_{qr}}$ \\ \hline 
    2 &  $\frac{\la qr\ra^2 \la r |\g^{\mu}|q]}{\sqrt{2}s_{qr}^2}$ &  $\frac{ [rq]^2 \la q |\g^{\mu}|r]}{\sqrt{2}s_{qr}^2}$ \\ \hline
     3 &  $\zeta r^{\mu}$ &  $\zeta r^{\mu}$ \\ \hline
      4 &  $\frac{\zeta[qr]r^{\mu}}{s_{qr}}$ &  $\frac{\zeta \la rq\ra r^{\mu}}{s_{qr}}$\\ \hline
       5 &  $\frac{\zeta[qr]\la q| \g^{\mu}| r]}{2s_{qr}}$ &  $\frac{\zeta \la rq\ra \la r| \g^{\mu}| q]}{2s_{qr}}$  \\ \hline
        6 &  $-\frac{\zeta q^{\mu}}{s_{qr}}$ &  $-\frac{\zeta q^{\mu}}{s_{qr}}$ \\ \hline
         7 &  $-\frac{\zeta \la q |\g^{\mu}|r]}{2s_{qr}}$ &  $-\frac{\zeta \la r |\g^{\mu}|q]}{2s_{qr}}$ \\ \hline
          8 &  $-\frac{\zeta q^{\mu}[qr]}{s_{qr}^2}$ &   $-\frac{\zeta q^{\mu}\la rq\ra}{s_{qr}^2}$ \\ \hline
\hline
    \hline
  \end{tabular}
    \caption{Definitions of the $\beta_{i}^{\mu}$ $\forall\,i=1\dotsm8$ with $\zeta=\frac{\sqrt{2}}{\sqrt{3}\tilde{m}}$ and $s_{qr}=-(q+r)^2$.} 
  \label{betas}
 \end{center}
\end{table}
Just by completeness we also express the 4 gravitino states $\tilde{\Psi}^{\mu}_{\lambda_p}(p)$ with $\lambda_p=++,--,+,-$, these take the following form:
\begin{align}
\bar{\tilde{\Psi}}_{++}(p)&=\beta_1^{*\mu}[r|+\beta_2^{*\mu}\la q|\tilde{m},\label{eq:gravitinobasis5}\\
\bar{\tilde{\Psi}}_{--}(p)&=-\beta_1^{\mu}\la r|+\beta_2^{\mu}[q|\tilde{m},\\
\bar{\tilde{\Psi}}_-^{\mu}(p)&=\beta_3^{*\mu}\la r|+(\beta_4^{*\mu}[q|+\beta_5^{*\mu}[r|)\tilde{m}+(\beta_6^{*\mu}\la r|+\beta_7^{*\mu}\la q|)\tilde{m}^2+\beta_8^{*\mu}[q|\tilde{m}^3,\\
\bar{\tilde{\Psi}}_+^{\mu}(p)&=\beta_3^{*\mu}[r|-(\beta_4^{\mu}\langle q|+\beta_5^{\mu}\langle r|)\tilde{m}+(\beta_6^{\mu}[r|+\beta_7^{\mu}[q|)\tilde{m}^2+\beta_8^{\mu}\langle q|\tilde{m}^3.\label{eq:gravitinobasis8}
\end{align}
Having the massive gravitino states in this kind of basis it is even more simple to handle the helicity amplitudes. For example we can verify that the gravitino states fulfill the normalization condition Eq.~(\ref{eq01:normalizationcondition}) i.e. taking $\lambda_p=-$, we have:
\begin{align}
\bar{\tilde{\Psi}}^{\mu}_-(p)\tilde{\Psi}_{\mu-}(p)&=\la rq\ra\big(\beta_3^{*\mu}\beta_{4\mu}+\beta^{*\mu}_{3}\beta_{8\mu}\tilde{m}^3+\beta_{6}^{*\mu}\beta_{4\mu}\tilde{m}^3+\beta_{6}^{*\mu}\beta_{8\mu}\tilde{m}^5-\beta_{7}^{*\mu}\beta_{5\mu}\tilde{m}^3\big)+\text{c.c.}\\
&=\la rq\ra\Big(-\frac{2\zeta^2[qr](r\cdot q)}{s_{rq}^2}\tilde{m}^3-\frac{\zeta^2[qr]}{2s_{qr}}\tilde{m}^3\Big)+\text{c.c.}\\
&=3\zeta^2\tilde{m}^3\\
&=2\tilde{m}^2,
\end{align}
as it can be noticed in the last calculation, the equations (\ref{eq:gravitinobasis1})-(\ref{eq:gravitinobasis4}) and (\ref{eq:gravitinobasis5})-(\ref{eq:gravitinobasis8}) are very convenient in order to handle huge and messy algebraic calculations.

\section{Helicity Amplitudes}\label{helicity_amplitudes}
\subsection{The goldstino equivalence theorem}

For a light gravitino it is possible to discuss its properties using the equivalence
theorem \cite{gge}, and replace the longitudinal components of the gravitino by the derivatives of the 
Goldstino field. For the strict massless case one can simply apply the massless helicity methods,
while for the massive case one needs to take into account the massive Dirac equation and the light-cone decomposition.
Considering the gravitino 4-momentun in spherical coordinates 
\begin{equation}\label{eq:gmomentum}
p^{\mu}=(E,|\vec{p}|\sin\theta\cos\phi,|\vec{p}|\sin\theta\sin\phi,|\vec{p}|\cos\theta) ,
\end{equation}
with $p^2=-\tilde{m}^2$. The polarization vectors take the following form
\begin{align}
\epsilon_{+}^{\mu}(p)&=\frac{1}{\sqrt{2}}(0,\cos\theta\cos\phi-i\sin\phi,\cos\theta\sin\phi+i\cos\phi,-\sin\theta),\\
\epsilon_{-}^{\mu}(p)&=-\frac{1}{\sqrt{2}}(0,\cos\theta\cos\phi+i\sin\phi,\cos\theta\sin\phi-i\cos\phi,-\sin\theta),\\
\epsilon_{0}^{\mu}(p)&=-\frac{1}{\tilde{m}}(|\vec{p}|,-E\sin\theta\cos\phi,-E\sin\theta\sin\phi,-E\cos\theta),
\end{align}
when on takes the limit $|\vec{p}|\to\infty$, one has $E\approx|\vec{p}|$, which implies that
\begin{align}
\epsilon_{\mu+}(p)p^{\mu}&=-\epsilon_{0+}(p)p^0+\vec{\epsilon}_+(p)\cdot\vec{p}\\
&=-\epsilon_{0+}(p)|\vec{p}|+|\vec{\epsilon}_+(p)||\vec{p}|\sin\theta,
\end{align}
 the condition  $p_{\mu}\epsilon^{\mu}_{\pm}(p)=0$ implies  $\epsilon_{+}^{\mu}(p)=0$ and  $\epsilon_{-}^{\mu}(p)=0$ when $|\vec{p}|\to\infty$. However  in this limit the polarization vector $\epsilon_{0}^{\mu}(p)$ has the following expression:
\begin{equation}
\epsilon_{\mu0}(p)=\frac{p_{\mu}}{\tilde{m}}.
\end{equation}
Thus the helicity states of the gravitino Eqs.~(\ref{eq:gravitinostate01})-(\ref{eq:gravitinostate04}) are reduced when one takes  into account  high energy limit, and now the surviving  gravitino states are only those of helicity $\pm1/2$, namely
\begin{align}\label{eq:gfappox1}
\tilde{\Psi}_{++}^{\mu}(p)&=0,\\\label{eq:gfappox2}
\tilde{\Psi}_{--}^{\mu}(p)&=0,\\\label{eq:gfappox3}
\tilde{\Psi}_{-}^{\mu}(p)&=\sqrt{\frac{2}{3}}\epsilon_{0}^{\mu}(p)u_-(p)=\sqrt{\frac{2}{3}}\left(\frac{p^{\mu}}{\tilde{m}}\right)u_-(p),\\\label{eq:gfappox4}
\tilde{\Psi}_{+}^{\mu}(p)&=\sqrt{\frac{2}{3}}\epsilon_{0}^{\mu}(p)u_+(p)=\sqrt{\frac{2}{3}}\left(\frac{p^{\mu}}{\tilde{m}}\right)u_+(p).
\end{align}
To convert then in to  coordinate space we need to replace $p^{\mu}\to i\partial^{\mu}$ in the gravitino field (\ref{eq:gfappox3})-(\ref{eq:gfappox4}) i.e. $\tilde{\Psi}_{\mu}(x)\to i\sqrt{\frac{2}{3}}\frac{\partial_{\mu}{\psi(x)}}{\tilde{m}}$, where $\psi(x)$ is the so-called spin-1/2 goldstino  state. After replacing the  gravitino field as goldstino approximation in the  lagrangian with gravitino  $\Psi^{\mu}(x)$  one obtain an effective lagrangian describing the interaction of the goldstino with chiral superfields, this  is given by \cite{takeo}:

\begin{equation}\label{eq:goldstinolagrangian}
 \mathcal{L}=\frac{i(m_{\phi}^2-m_{\chi}^2)}{\sqrt{3}\tilde{m}M}(\bar{\psi}\chi_R)\phi^*-\frac{im_{\lambda}}{8\sqrt{6}\tilde{m}M}\bar{\psi}[\gamma^{\mu},\gamma^{\nu}]\lambda^{(a)}F^{(a)}_{\mu\nu}+h.c.
\end{equation}

In this approximation when one assemble the HA's from the Feynman rules, the goldstino field is just a Dirac spinor that is well known in literature.


\subsection{Massless and massive gauge boson amplitudes}
It is known that tree-level amplitudes that include $n$ massless  gauge bosons of configurations $(+,+,\dotsm,+)$ or $(-,-,\dotsm,-)$ vanish exactly; one needs to have at least two helicities of each sign in order to have a non-vanishing amplitude, i.e. $(-,-,+,+,\dotsm,+)$ or $(+,+,-,-,\dotsm,-)$. This result also extends to amplitudes involving massless gravitons.\\
Now, it is the case that we are interested in evaluating processes involving massive gauge bosons, as one of the goals of LHC is to probe the mechanism of EWSB. Thus, it should be interesting to discuss to what extend the results of massless gauge boson scattering generalize to the massive case. In fact, this has addressed in [selection rules paper], with the finding that certain vanishing amplitudes for the massless case i.e. with  $(-,-,+,+,\dotsm,+)$  or  $(+,+,-,-,\dotsm,-)$ configuration become non-vanishing but with factors of the form $\mathcal{O}(\frac{m}{E})$, with $m$ being the gauge boson mass and $E$ denoting the C.M. energy of the physical process under consideration. This result could also be understood by relying on the equivalence theorem, namely when the vector boson scattering is approximated by the pseudo-goldstone bosons.

\subsection{Amplitudes for massive gravitinos}

A similar result is expected to hold for the massive gravitino scattering. Namely, some amplitudes with some helicity configurations that vanish in the massless case would get corrections of the form $\mathcal{O}(\frac{\tilde{m}}{E})$, where $\tilde{m}$ denotes the gravitino mass, and $E$ is the typical energy of the physical process. \\
Again this can be induced by relying on the SUGRA equivalence theorem, where the $\pm$ helicity states associated with the goldstino that arises from the Super-Higgs mechanism, that it is used to break supersymmetry and induce masses of the superpartners, including the gravitino.\\
As we are not aware  of the corresponding discussions on the general case, we shall look at some specific process in order to identify the corresponding results, namely to lock at the result for the MHV amplitude and try to identify the possible corrections, then to identify such non-MHV amplitudes that become non-vanishing and which amplitudes remain vanishing.

\section{A comparison of amplitudes with gravitino and goldstino}\label{comparison_amplitudes}
\subsection{The 2-body neutralino decay $\tilde{\chi}_0\to\tilde{\Psi}^{\mu}\,\gamma$ with LSP gravitino in the final state}
One of the simplest process that allows us to study the helicity configurations for scattering amplitudes is  the 2-body decay of a neutralino into a gravitino and a photon. Using the interactions of the MSSM with gravity ($\tilde{\chi}_0\to\tilde{G}\gamma$ ) \cite{ellis2} we write the amplitude for the simple Feynman diagram Fig.~(\ref{fig:neutralinodecay}) that contributes is as follows:
\begin{align}\label{eq:001:ampneutralino}
\mathcal{M}_{\lambda_q\lambda_p\lambda_k}^{c}&=\frac{1}{4M}C_{\chi\gamma}\bar{\tilde{\Psi}}_{\mu\lambda_p}(p)\left(k_{\nu}[\gamma^{\nu},\gamma^{\sigma}]\gamma^{\mu}\right)\epsilon_{\sigma\lambda_k}(k)u_{\lambda_q}(q)\\
&=\frac{1}{4M}C_{\chi\gamma}\bar{\tilde{\Psi}}_{\mu\lambda_p}(p)\big(\epsilon^{\mu}_{\lambda_k}(k)\s{k}-k^{\mu}\s{\epsilon}_{\lambda_k}(k)\big)u_{\lambda_q}(q)\\
&=\frac{1}{4M}C_{\chi\gamma}\bar{\tilde{\Psi}}_{\mu\lambda_p}(p)X^{\mu}_{\lambda_k}(k)u_{\lambda_q}(q),
\label{eq:002:ampneutralino}
\end{align}
with  $C_{\chi\gamma}=U_{i1}\cos\theta_W+U_{i2}\sin\theta_W$. The momenta assignments for this decay is $p$ for the gravitino field ($\Psi_{\mu}(p)$), $q$ for the Neutralino ($\tilde{\chi}_0(q)$) and $k$ for the photon ($\gamma(k)$) and $\lambda_{q},\,\lambda_{p}$ and $\lambda_{k}$ are their helicities labels, we have also defined in Eq.~(\ref{eq:002:ampneutralino}) $X^{\mu}_{\lambda_k}(k)=\epsilon^{\mu}_{\lambda_k}(k)\s{k}-k^{\mu}\s{\epsilon}_{\lambda_k}(k)$. 
 There are 16  helicity amplitudes to compute, but by complex conjugate symmetry we just need to calculate half of them. 

\vspace{0.03\linewidth}

\begin{minipage}{\linewidth}
\begin{figure}[H]
\centering
\begin{picture}(-10,75)
\put(-70,0){\includegraphics[scale=0.6]{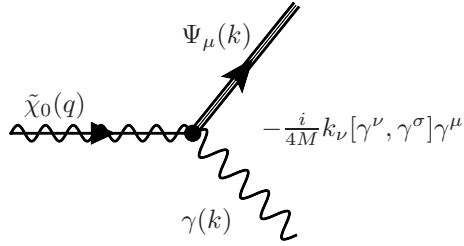}}
\put(-64,49){$\tilde{\chi}_0(q)$}
\put(-5,5){$\gamma(k)$}
\put(-5,75){$\Psi_{\mu}(k)$}
\put(25,40){$-\frac{i}{4M}k_{\nu}[\g^{\nu},\g^{\sigma}]\gamma^{\mu}$}
\end{picture}
\caption{Feynman diagram for gravitino interaction with neutralino and photon}
        \label{fig:neutralinodecay}
\end{figure}
      \end{minipage}
     
       \vspace{0.03\linewidth}
The nonzero HA are shown in  Table (\ref{table8})
\begin{table}[H]
\begin{center}
  \begin{tabular}{  || c | c |c || }
    \hline \hline
    $\lambda_q,\,\lambda_p,\,\lambda_k$ & $\mathcal{M}_{\lambda_q,\,\lambda_p,\,\lambda_k}^c$ & $\mathcal{M}_{\lambda_q,\,\lambda_p,\,\lambda_k}^a$\\ \hline
   \hline\hline
     $-,++,+$ & $ \frac{C_{\chi\gamma}[ r_2q_2]^2}{M\langle r_1r_2\rangle}m_{\tilde{\chi}_0}$ & $0$\\ \hline 
    $-,-,-$ & $\frac{C_{\chi\gamma}s_{r_2q_2}}{\sqrt{3}\tilde{m}[r_2q_2]}\langle r_2q_2\rangle[r_2r_1]$ &  $\frac{C_{\chi\gamma}s_{r_2q_2}}{\sqrt{3}\tilde{m}[r_2q_2]}\langle r_2q_2\rangle[r_2r_1]$ \\ \hline
\hline
    \hline
  \end{tabular}
    \caption{Helicity Amplitudes for the 2-body Neutralino decay $\chi_{0}\to\gamma\,G$. Here $\mathcal{M}_{\lambda_q,\,\lambda_p,\,\lambda_k}^c$ represent the helicity amplitudes complete, this means for the massive spin-3/2 gravitino and, $\mathcal{M}_{\lambda_q,\,\lambda_p,\,\lambda_k}^a$ are  for the approximation of the gravitino to goldstino. }
  \label{table8}
 \end{center}
\end{table}
It is quite remarkable that the ``massless" approximation for the helicity amplitudes with gravitino is exactly the helicity amplitude for the goldstino with just one configuration of helicities, this is $\mathcal{M}_{-,-,-}^c\equiv\mathcal{M}^a_{-,-,-}$.

The squared and averaged amplitude take the form
\begin{align}
\langle|\mathcal{M}|^2\rangle&=\frac{C_{\chi\gamma}^2}{2M^2}\big(2|\mathcal{M}_{-,++,+}|^2|+2\mathcal{M}_{-,-,-}|^2\big)\\
&=\frac{C_{\chi\gamma}^2}{M^2}\left(\frac{s_{q_2r_2}^2m_{\tilde{\chi}_0}^2}{s_{r_1r_2}}+\frac{s_{r_2q_2}^2}{3\tilde{m}^2}s_{r_2r_1}\right)\\
&=\frac{C_{\chi\gamma}^2}{M^2}\left(\frac{(m_{\tilde{\chi}_0}^2-\tilde{m}^2)^2}{3\tilde{m}^2}(3\tilde{m}^2+m_{\tilde{\chi}_0}^2)\right)\\
&=\frac{C_{\chi\gamma}^2m_{\tilde{\chi}_0}^6}{M^2}\left(1-\frac{\tilde{m}^2}{m_{\tilde{\chi}_0}^2}\right)^2\left(\frac{1}{3}+\frac{\tilde{m}^2}{m_{\tilde{\chi}_0}^2}\right),
\end{align}
then the decay width $\Gamma$ for the 2-body neutralino decay ($\tilde{\chi}_0\to\gamma\tilde{G}$) is as follows
\begin{align}
\Gamma_{\tilde{\chi}_0\to\gamma\tilde{G}}
&=\frac{C_{\chi\gamma}^2m_{\tilde{\chi}_0}^5}{16\pi M^2\tilde{m}^2}\left(1-\frac{\tilde{m}^2}{m_{\tilde{\chi}_0}^2}\right)^3\left(\frac{1}{3}+\frac{\tilde{m}^2}{m_{\tilde{\chi}_0}^2}\right),
\end{align}
where $s_{r_1r_2}=m_{\tilde{\chi}_0}^2$, $s_{r_2q_2}=m_{\tilde{\chi}_0}^2-\tilde{m}^2$ and  $s_{q_2r_1}=0$. 
\subsection{Production of light gravitino at colliders: $e^{+}e^{-}\to\tilde{G}\tilde{G}$}
 We will compute the scattering amplitude for the reaction $e^{-}e^{+}\to\,G\,G$  with the gravitino approximation to goldstino (massless) \cite{monophoton}. Each Feynman diagram of Fig.~(\ref{fig:lightgravitino}) contributes to the total amplitude:
 \begin{equation}
 \mathcal{M}=\mathcal{M}^c+\mathcal{M}^u+\mathcal{M}^t
 \end{equation}      
where
\begin{align}
\mathcal{M}^{c}&=-\frac{m_{\tilde{e}_{\lambda_1}}^{2}}{F^2}(\mathcal{T}^{t}-\mathcal{T}^u),\\
\mathcal{M}^t&=-\frac{m_{\tilde{e}_{\lambda_1}}^{4}}{F^2(t-m_{\tilde{e}_{\lambda_1}}^2)}\mathcal{T}^{t},\\
\mathcal{M}^u &=\frac{m_{\tilde{e}_{\lambda_1}}^{4}}{F^2(u-m_{\tilde{e}_{\lambda_1}}^2)}\mathcal{T}^{u},
\end{align}
where $m_{\tilde{e}_{\lambda_1}}$ is the  selectron masses, $\lambda_1=\pm$, 

\vspace{0.03\linewidth}

\begin{minipage}{\linewidth}
\begin{figure}[H]
\centering
\begin{picture}(150,75)
\put(-70,0){\includegraphics[scale=0.6]{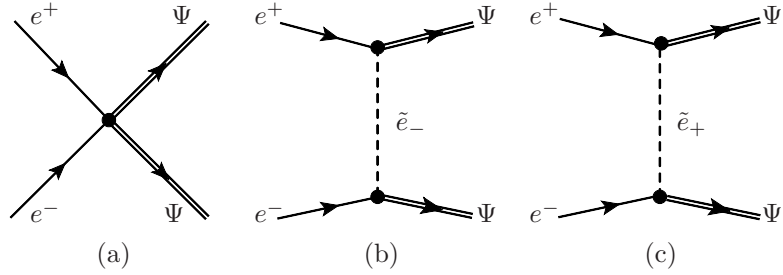}}
\put(-62,75){$e^{+}$}
\put(-62,0){$e^{-}$}
\put(-12,0){$\Psi$}
\put(-9,75){$\Psi$}
\put(-37,-15){(a)}
\put(75,35){$\tilde{e}_{-}$}
\put(22,75){$e^{+}$}
\put(22,0){$e^{-}$}
\put(105,0){$\Psi$}
\put(105,75){$\Psi$}
\put(63,-15){(b)}
\put(180,35){$\tilde{e}_{+}$}
\put(125,75){$e^{+}$}
\put(125,0){$e^{-}$}
\put(212,0){$\Psi$}
\put(212,75){$\Psi$}
\put(168,-15){(c)}
\end{picture}
 \vspace{0.02\linewidth}
\caption{Feynman diagrams for gravitino production at $e^{+}e^{-}$ colliders}
        \label{fig:lightgravitino}
\end{figure}
      \end{minipage}
     
       \vspace{0.03\linewidth}

\begin{align}\label{eq:0:0:09}
\mathcal{M}^{c}_{-,+} &=-\frac{m_{\tilde{e}_{-}}^2}{F^2}(\mathcal{T}^t_{-,+}-\mathcal{T}^u_{-,+})=-\frac{m_{\tilde{e}_{-}}^2}{F^2}[31]\la24\ra\\
\mathcal{M}^{u}_{-,-} &=\frac{m_{\tilde{e}_-}^4}{F^2(u-m_{\tilde{e}_-}^2)}[41]\la 23\ra\\
\mathcal{M}^{t}_{-,+} &=-\frac{m_{\tilde{e}_-}^4}{F^2(t-m_{\tilde{e}_-}^2)}[31]\la 24\ra
\end{align}
The nonzero helicity amplitudes are shown in  Table (\ref{table9})
\begin{table}[H]
\begin{center}
  \begin{tabular}{  || c | c || }
    \hline \hline
    $\lambda_1\lambda_2\lambda_3\lambda_4$ & $\mathcal{M}_{\lambda_1\lambda_2\lambda_3\lambda_4}$ \\ \hline
   \hline
     $-,+,+,-$ & $-\frac{m_{\tilde{e}_-}^2t}{F^2(t-m_{\tilde{e}_-}^2)}[31]\la 24\ra $\\ \hline
    $-,+,-,+$ & $\frac{m_{\tilde{e}_-}^2u}{F^2(u-m_{\tilde{e}_-}^2)}[41]\la 23\ra $\\ \hline
\hline
  \end{tabular}
    \caption{Helicity Amplitudes for the reaction $e^{-}e^{+}\to\tilde{G}\tilde{G}$}
  \label{table9}
 \end{center}
\end{table}
\section{Conclusions}\label{conclusions}
In this paper we have developed the formal tools needed to construct helicity amplitudes for massive gravitino in $\mathcal{N}=1$ SUGRA. 
We started by considering the helicity states for massive spin-3/2 particles, 
which involves the solutions of Rarita-Schwinger equation. Adopting the helicity bra-ket notation for these solutions, they were expressed in a convenient way for assembling helicity amplitudes  and were used  to derive the interactions of the gravitino with matter and gauge fields within $\mathcal{N}=1$ Supergravity. 
We also have studied  the corresponding interactions of  goldstinos, relying on the goldstino-gravitino equivalence theorem. 
Finally to appreciate the power of the method we have evaluated the cross sections for
$e^{+}\,e^{-}\to GG$ and the width decay $\tilde{\chi}_0\to Z \tilde{\psi}_{\mu}$.  It was shown in Tables \ref{table8} and \ref{table9} how the Spinor Helicity Formalism (massless and massive cases) reduce the expressions for scattering amplitudes, allowing to calculate effectively the physical observables. 
\section*{Acknowledgments}
We would like to acknowledge the support of CONACYT and SNI. Bryan Larios is supported by a CONACYT graduate student fellowship.

\appendix
\section{Basics of the massless helicity formalism}
In this appendix it is intruduced  the properties for the massless spinors that are use through this paper, most of them were taking from Ref.~\cite{srednicki}.  

Using the powerful spinor bra-ket notation, the 4-component Dirac spinor are rewritten as  follows  
\begin{align}\label{eq:bknotation}
u_-(p)&=v_{+}(p)=|p],\\
u_+(p)&=v_-(p)=|p\rangle,\\
\bar{u}_+(p)&=\bar{v}_-(p)=[p|,\\
\bar{u}_-(p)&=\bar{v}_+(p)=\langle p|,
\end{align}
which obey the relations
\begin{align}
u_{s}(p)\bar{u}_s(p)&=\frac{1}{2}(1+s\g_5)(-\s{p})\\
v_{s}(p)\bar{v}_s(p)&=\frac{1}{2}(1-s\g_5)(-\s{p})
\end{align}
where $s=\pm$ indicates the helicity. Spinor products are antisymmetric
\begin{align}
	\bar{u}_+(p)u_-(k)&=[pk] =-[kp]=-\bar{u}_+(k)u_-(p), \\
	\bar{u}_-(p)u_+(k)&=\langle pk\rangle=-\langle kp\rangle=\bar{u}_-(k)u_+(p) ,
\end{align}
taking the last results into account one also have that the spinor product fulfill  $[qq]=\la qq\ra=0$, the type of spinor products $[k p \ra$ and $\la pk]$ are also null.  

For real momenta these spinor products satisfy
\begin{align}
	\langle pk\rangle&=[kp]^\ast,\\
	[kp]&=\langle pk\rangle^*,\\
[pq]\langle pq\rangle&=s_{pq}=-(p+q)^2=-2p\cdot q .
\end{align}
Other useful properties are the following
\begin{align}
	[k|\gamma^\mu|p\rangle&=\langle p|\gamma^\mu|k] , \label{sq_br_1} \\
	[k|\gamma^\mu|p\rangle^\ast&=[p|\gamma^\mu|k\rangle  \label{sq_br_2} \\
	\langle p|\slashed{k}|q]&=-\langle pk\rangle[kq],\\
	 	\langle p|\gamma^\mu|p]&=2p^\mu, .
	 \end{align}
Fierz identity is also a useful property, this take the following form 
\begin{equation}
	\langle p|\gamma^\mu|q]\langle r|\gamma_\mu|w]=2\langle pr\rangle[qw] .
\end{equation}
From the completeness relation, one is able to express $\s{p}$ as a  product of spinors
\begin{equation}
\s{p}=-(|p]\la p|+|p\ra[p|)
\end{equation}




\begin{thebibliography}{9}


\bibitem{PT} S.~J.~Parke and T.~R.~Taylor, \textit{Phys. Rev. Lett.} {\bf 56}, 2459 (1986).

\bibitem{witten}  E. Witten, \textit{Commun. Math. Phys}. \textbf{252}, 189 (2004) [arXiv:hep-th/0312171].

\bibitem{BCJ} Z.~Bern,~J.~J.~Carrasco, and H.~Johansson, \textit{Phys. Rev. D} {\bf 78}, 0805011 (2008) [arXiv:0805.3993 [hep-ph]].

\bibitem{dixon1} L.J.~Dixon, [arXiv: 1310.5353 [ hep-th]].

\bibitem{dixon2}  L.~J.~Dixon, \textit{J. Phys. A}  {\bf 44}, 454001 (2011) [arXiv: 1105.0771[ hep-th]].


\bibitem{Nima1} N.~Arkani-Hamed and J.~Kaplan, \textit{JHEP} {\bf 0804}, 076 (2008) [arXiv:0801.2385 [hep-th]]. 


\bibitem{Nima2} N.~Arkani-Hamed, F~Cachazo, C.~Cheung, and J.~Kaplan, \textit{JHEP} {\bf 1003}, 020 (2010) [arXiv:0907.5418 [hep-th]]. 


\bibitem{Nima4} N.~Arkani-Hamed, J.~L.~Bourjaily, F~Cachazo, S.~Caron-Huot, J.~Trnka, \textit{JHEP} {\bf 1101}, 041 (2011) [arXiv:1008.2958 [hep-th]].

\bibitem{Nima5} N.~Arkani-Hamed, J.~L.~Bourjaily, F~Cachazo, et al., [arXiv:1212.5605 [hep-th]].

\bibitem{KLT} H.~Kawai, D.~C.~Lewellen, and S.~H.~H.~Tye, \textit{Nucl. Phys. B} {\bf 269}, 1 (1986).



\bibitem{BCF} R.~Britto, F.~Cachazo, and B.~Feng, \textit{Nucl. Phys. B} {\bf715}, 499 (2005) [hep-th/0412308]. 

\bibitem{BCFW} R.~Britto, F.~Cachazo, B.~Feng, E.~Witten, \textit{Phys. Rev. Lett. } {\bf 94}, 181602 (2005) [hep-th/0501052]. 

\bibitem{Nima3} N.~Arkani-Hamed, F~Cachazo, and J.~Kaplan, \textit{JHEP} {\bf 1009}, 016 (2010) [arXiv:0808.1446 [hep-th]]. 

\bibitem{le}	J.~L.~D\'iaz-Cruz, John Ellis, Keith A.~ Olive, Yudi Santoso, \textit{JHEP} {\bf0705}, 003, 2007, [arXiv:hep-ph/0701229v1].





\bibitem{gge} R.~Casalbuoni, S.~De Curtis, D.~Dominici, F.~Feruglio, and R. Gatto, \textit{Physics Letters B} {\bf 215},  313-316, 1988.
\bibitem{fayet1}  P.~Fayet, \textit{Phys.Lett.} \textbf{70B} (1977) 461.

\bibitem{fayet2} P.~Fayet, \textit{Phys. Lett. B.} \textbf{175} (1986) 471.

\bibitem{fayet3} P.~Fayet, \textit{Phys. Lett. B.} \textbf{84} (1979) 421.

\bibitem{fayet4} P.~Fayet, \textit{Phys. Lett. B.} \textbf{86} (1979) 272.
 
 \bibitem{fayet5} P.~Fayet, \textit{Conference Proc.}  LPTENS-81-9 (1981) 347.
\bibitem{us1} J.~Lorenzo D\'iaz-Cruz, B.~Larios, \textit{Eur. Phys. J. C} (2016) {\bf 76}, 157, [arXiv:1510.01447v2 [hep-ph] ].




 \bibitem{novaes} S.~F.~Novaes and, D.~Spehler, \textit{Nucl. Phys. B} \textbf{371} (1992) 618-636;\\
Luis A. Anchordoqui, Ignatios Antoniadis, De-Chang Dai, Wan-Zhe Feng, Haim Goldberg, Xing Huang, Dieter Lust, Dejan Stojkovic, Tomasz R. Taylor, \textit{Phys. Rev. D} \textbf{90}, 066013 (2014), [	arXiv:1407.8120 [hep-ph]].

 \bibitem{roy1} T.~Bhattacharya, P.~Roy, \textit{Nuclear Physics B}
 {\bf328}, 469-480,  1989.

  
  
 \bibitem{roy2} T.~Bhattacharya, P.~Roy, \textit{Nuclear Physics B}
 {\bf328}, 481-498,  1989.  

 
 

\bibitem{madgraph} Johan Alwall, Michel Herquet, Fabio Maltoni, Olivier Mattelaer, Tim Stelzer, \textit{High Energ. Phys.} (2011) 2011: \textbf{128},   [arXiv:1106.0522v1 [hep-ph]].

\bibitem{schwartz} M.~D.~Schwarz, \textit{Quantum Field Theory and the Standard Model}, Cambridge University Press, 2014.
\textbf{252} 189-258 (2004), [arXiv:hep-th/0312171].






\bibitem{srednicki} M.~Srednicki, \textit{Quantum Field Theory}, Cambridge University Press, 2007.


\bibitem{elvang} H.~Elvang and Y.~Huang, \textit{Scattering Amplitudes in Gauge Theory and Gravity}, Cambridge University Press, 2015, [arXiv:1308.1697v2 [hep-th]].




\bibitem{lorenzo} J.~L.~D\'iaz-Cruz, B.~Larios, O.~Meza and, J.~ Reyes, \textit{Rev. Mex. Fis. E} \textbf{61(2)} (2015) 104. English version: [arXiv:1511.07477 [physics.gen-ph]].








\bibitem{boels} R.~Boels, \textit{JHEP} \textbf{1001} 010 (2010), [arXiv:0908.0738 [hep-th]].

\bibitem{weinzierl} C.~Schwinn and S.~Weinzierl, \textit{JHEP} \textbf{0704} (2007) 072, [arXiv:hep-ph/0703021 [hep-ph]].


\bibitem{spinorsextras} J.~Kuczmarski, [arXiv:1406.5612 [hep-ph]].



\bibitem{BOL} J.~Lorenzo D\'iaz-Cruz, B.~Larios, O.~Meza, \textit{Journal of Physics Conference Series} {\bf761} (1), 2016, [arXiv:1608.04129v1 [hep-ph]].

\bibitem{rarita} W.~Rarita and J.~Schwinger, \textit{Phys. Rev.} \textbf{60} (1941) 60.
\bibitem{auvil} P.~R.~Auvil and J.~J.~Brehm, \textit{Phys. Rev. }145 (1966) \textbf{1152}.
\bibitem{takeo} T.~Moroi, [arXiv:hep-ph/9503210].
\bibitem{dittmaier} S. Dittmaier, Phys. Rev. D \textbf{59} 016007 (1998), [hep-ph/9805445].
 \bibitem{byo} B.~Larios, O.~Meza, work in progress.
\bibitem{ellis2} J.~Ellis, , K.~A.~Olive, Y.~Santoso, V.~C.~Spanos, \textit{Phys. Lett. B} \textbf{588} (2004), [arXiv:hep-ph/0312262v4].
\bibitem{monophoton} K.~Mawatari, B. Oexl, \textit{Eur.Phys.J.} \textbf{C74} (2014) no.6, 2909, [ arXiv:1402.3223 [hep-ph]].































\end{thebibliography}
\end{document}